\input harvmac
%\draft
\noblackbox
%-------------------------
% This paper uses harvmac
%-------------------------
\overfullrule=0pt
\def\Title#1#2{\rightline{#1}\ifx\answ\bigans\nopagenumbers\pageno0\vskip1in
\else\pageno1\vskip.8in\fi \centerline{\titlefont #2}\vskip .5in}

%
%-------------------
%  definitions
%
\def\RN{Reissner-Nordstr\"om}
\font\cmss=cmss10 \font\cmsss=cmss10 at 7pt
\def\IZ{\relax\ifmmode\mathchoice
   {\hbox{\cmss Z\kern-.4em Z}}{\hbox{\cmss Z\kern-.4em Z}}
   {\lower.9pt\hbox{\cmsss Z\kern-.4em Z}}
   {\lower1.2pt\hbox{\cmsss Z\kern-.4em Z}}\else{\cmss Z\kern-.4emZ}\fi}

\def\sbh{S_{BH}}
%
%-------------------
% references
%
\lref\gm{G. Horowitz and D. Marolf, hep-th/9605224.}
\lref\ms{J. Maldacena and L. Susskind, hep-th/9604042.}
\lref\bl{V. Balasubramanian and F. Larsen, hep-th/9604189.}
\lref\jmn{J. Maldacena, hep-th/9605016.}
\lref\jkm{C. Johnson, R. Khuri and R. Myers, hep-th/9603061.}
\lref\kt{I. Klebanov and A. Tseytlin, hep-th/9604089.}
\lref\dvv{R. Dijkgraaf, E. Verlinde and H. Verlinde, hep-th/9603126.}
\lref\hlm{G. Horowitz, D. Lowe and J. Maldacena, hep-th/9603195.}
\lref\ms{J. Maldacena and A. Strominger, hep-th/9603060.}
\lref\hms{G. Horowitz, J. Maldacena and A. Strominger, hep-th/9603109.}
\lref\sch{J. Schwarz, Phys. Lett. {\bf B272} (1991) 239.}
\lref\wol{E. Witten and D. Olive, Phys. Lett. {\bf B78} (1978) 97.}
\lref\swone{N. Seiberg and E. Witten, hep-th/9407087, Nucl. Phys. {\bf
B426}
(1994) 19. }
\lref\cpw{S.Coleman, J. Preskill and  F. Wilczek,  Phys. Rev. Lett. 
{\bf 67} (1991) 1975; Nucl. Phys. {B378} (1992) 175.}
\lref\fg{S. Giddings, J. Harvey, J. Polchinski, S. Shenker and A.
Strominger,
hep-th/9309152, Phys. Rev. {\bf D50} (1994) 6422.}
\lref\rk{S. Ferrara and R. Kallosh, hep-th/9603}
\lref\gkp{S. Gubser, I. Klebanov and A. Peet, hep-th/9602135.}
\lref\dbr{J. Polchinski, S. Chaudhuri, and C. Johnson, hep-th/9602052.}
\lref\jp{J. Polchinski, hep-th/9510017.}
\lref\dm{S. Das and S. Mathur, hep-th/9601152.}
\lref\witb{E. Witten, hep-th/9510135.}
\lref\ghas{G. Horowitz and A. Strominger, hep-th/9602051.}
\lref\polc{J. Dai, R. Leigh and J. Polchinski, Mod. Phys.
Lett. {\bf A4} (1989) 2073.}
\lref\ascv{A. Strominger and C. Vafa, hep-th/9601029.}
\lref\hrva{P. Horava, Phys. Lett. {\bf B231} (1989) 251.}
\lref\cakl{C. Callan and I. Klebanov, hep-th/9511173.}
\lref\prskll{J. Preskill, P. Schwarz, A. Shapere, S. Trivedi and
F. Wilczek, Mod. Phys. Lett. {\bf A6} (1991) 2353. }
\lref\sbg{S. Giddings, Phys. Rev {\bf D49} (1994) 4078.}
\lref\bhole{G. Horowitz and A. Strominger,
Nucl. Phys. {\bf B360} (1991) 197.}
\lref\bekb{J. Bekenstein, Phys. Rev {\bf D12} (1975) 3077.}
\lref\hawkb{S. Hawking, Phys. Rev {\bf D13} (1976) 191.}
\lref\wilc{P. Kraus and F. Wilczek, hep-th/9411219, Nucl. Phys.
{\bf B433} (1995) 403. }
\lref\stas{A.~Strominger and S.~Trivedi,  Phys.~Rev. {\bf D48}
 (1993) 5778.}
\lref\lawi{F. Larsen and F. Wilczek, hep-th/9511064.}
\lref\bek{J. Bekenstein, Lett. Nuov. Cimento {\bf 4} (1972) 737,
Phys. Rev. {\bf D7} (1973) 2333, Phys. Rev. {\bf D9} (1974) 3292.}
\lref\hawk{S. Hawking, Nature {\bf 248} (1974) 30, Comm. Math. Phys.
{\bf 43} (1975) 199.}
\lref\cama{C. Callan and J. Maldacena, hep-th/9602043.}
\lref\spn{J. Breckenridge, R. Myers, A. Peet and C. Vafa, hep-th/9602065.}
\lref\vbd{J. Breckenridge, D. Lowe, R. Myers, A. Peet, A. Strominger 
and C. Vafa, hep-th/9603078.}
\lref\hpc{S. Hawking, private communication.}
\lref\qhair{M. Bowick, S, Giddings, J. Harvey, G. Horowitz and A. Strominger,
    Phys. Rev. Lett. {\bf 61} (1988) 2823.}
\lref\gpp{S. Giddings, J. Polchinski and J. Preskill, unpublished (1990).}
\lref\kw{L. Krauss and F. Wilczek, Phys. Rev. Lett. {\bf 62} (1989) 1221.}
%-------------------
% title page
%-------------------
%
\Title{\vbox{\baselineskip12pt
\hbox{hep-th/9606016}\hbox{RU-96-47}}}
{\vbox{
\centerline {Statistical Hair on } 
\centerline{Black Holes}  }}
\centerline{Andrew Strominger}
\vskip.1in
\centerline{\it Department of Physics and Astronomy, Rutgers University,
Piscataway, NJ 08855}
\centerline{\it and}
\centerline{\it Department of Physics, University of California,
Santa Barbara, CA 93106}

\vskip.1in
\vskip1in
\centerline{\bf Abstract}
The Bekenstein-Hawking entropy for certain BPS-saturated black 
holes in string theory has recently been derived by counting 
internal black hole microstates at weak coupling. We argue 
that the black hole microstate can be measured by interference 
experiments even in the strong coupling region where there is 
clearly an event horizon. Extracting information which is naively 
behind the event horizon is possible due to the existence of 
statistical quantum hair carried by the black hole. This quantum 
hair arises from the arbitrarily large number 
of discrete gauge symmetries present in string theory.

\Date{}
%
%----------------------
% Body of Paper
%----------------------

More than twenty years ago Bekenstein \bek\ and Hawking \hawk\ 
showed that black holes possess a macroscopic 
thermodynamic entropy equal to one quarter the area of the event horizon 
in Planck units: $\sbh =A/4$. This beautiful result cried out for a microscopic 
statistical derivation. However, because the Planck length appears in the relation it is likely that a quantum theory of gravity is required for such 
a derivation.  
In the last several months it has been found that string theory 
- a candidate for a quantum theory of gravity -
can provide a precise derivation of this entropy in a 
variety of circumstances \refs{\ascv \dm \cama \ghas \spn \vbd 
\gkp \dvv \ms \jkm \kt \hms \hlm \jmn \bl \ms -\gm}.

The derivation exploited the existence of 
two distinct descriptions of extremal BPS- saturated states in $N=4$ or 
$N=8$ supersymmetric string theories.  
The first description is as a semiclassical quantum state 
corresponding to the usual extremal 
\RN\ solutions\foot{Or generalizations thereof involving extra 
scalar fields or more dimensions, depending on the context.}. 
The second description is as a quantum bound state of elementary 
D-brane solitons \refs{\polc \hrva -\jp} and strings. The logarithm of the bound state degeneracy
(as a function of the charges) as 
computed in the second picture agrees (for large charges) 
with $\sbh$ as computed in the first.

These two pictures are 
physically relevant at different values of the string coupling
$g_s$ \ascv. At 
strong coupling\foot{By which we mean $g_s \sim 1$. For $g_s>>1$ the 
theory may have a weakly coupled dual description.}, 
string perturbation theory is 
divergent, fluctuations in the D-brane bound state configuration are large, 
and the D-brane description is inaccurate.  
Supersymmetric nonrenormalization theorems nevertheless 
protect the low energy effective action. 
Hence the black hole solution, which is independent of $g_s$,  
should still provide an accurate description of the state. 
In a slight abuse of terminology we refer to this 
region of couplings as the 
`black hole phase' (we do not mean to suggest there is a
sharp phase transition at some value of $g_s$). 
For very weak string coupling, string 
perturbation theory is 
good, and the D-brane picture is valid. However in this region 
the string length,
which grows (in Planck units) as the coupling decreases, becomes 
larger than the 
Schwarzchild radius. The black hole is smaller than a typical 
string. At scales shorter 
than the string length string theory gives drastic modifications of our 
notion of 
Riemmanian geometry, even at the classical level. In particular 
there is no obvious notion of causality or 
an event horizon at such short scales.
Hence the black hole picture is inaccurate in this 
`D-brane phase'. 

There is no overlapping range of validity of the two pictures (at least in 
examples considered so far). 
The agreement between the entropies as computed in the two pictures was 
nevertheless expected because, in an $N=4$ or $N=8$ supersymmetric theory the 
number of BPS states typically does 
not change as a function of the coupling \wol.
This fact 
enables one to extrapolate the D-brane phase calculation into the black hole 
phase. Such extrapolations will certainly not be possible for general 
S-matrix elements in the theory, which will depend strongly on the coupling. 

One would like to apply the new insight from string theory 
into black hole entropy 
to the black hole information 
puzzle. One form of the argument that information is lost 
is briefly as follows. Throw a neutral 
string in a definite 
quantum state into a large extremal black hole. The black hole then becomes 
nonextremal and acquires a nonzero Hawking temperature. It reradiates back to 
extremality via Hawking emission. In the semiclassical regime the outgoing 
quanta arises from a pair production process outside the horizon, and 
hence is blind 
to the internal quantum state of the black hole. Hence the final state 
of extremal black hole + outgoing quanta is insensitive to some aspects 
of the initial state. Unitarity is violated 
and information is lost. Further discussions of 
this example, including subtleties which could invalidate the conclusion that 
information is lost, can be found in \prskll, \stas.

This puzzle exists only in the black hole phase. In the D-brane 
phase there is 
no event horizon and hence no reason to expect that the outgoing 
quanta cannot access information about the incident quanta. Indeed the 
scattering is explicitly calculable and perturbatively unitary. 
One point of 
view \hpc, is that the system is like a neutron star. Everyone would 
agree that 
the quantum state of a neutron star can be measured in principle. 
However if the 
gravitational coupling is turned up, the neutron star will collapse into a 
black hole at a critical value of the coupling. 
Then by the no-hair theorem it would appear that no information 
about the quantum state is available. Hence the fact that there is 
perturbatively unitary scattering in the D-brane phase does not 
immediately imply that there is no 
unitarity violation in the black hole phase: there could be 
some kind of collapse at 
a critical value of $g_s$.

Clearly in order to address the information puzzle we must analyze the
black hole phase.
Herein we address the simplest question: Can the quantum state of an extremal 
black hole be measured\foot{As with any single quantum object, the 
state may change during the measurement process and we 
can at most hope to determine the quantum state after the  
measurement is completed.} in the black hole phase? In the 
D-brane phase the state can clearly be measured by scattering strings
off of the bound state.
However we do not know how to compute string scattering in the black 
hole phase, and the semiclassical no-hair theorem 
suggests that the scattering
depends only on a few charges and is 
insensitive to the specific black hole microstate.

A simple argument shows that the answer to this question is
nevertheless yes, assuming the existence of a 
low-energy effective field theory on 
scales large compared to the size of the black hole\foot{The 
reader may object that this amounts to assuming the answer. 
If this assumption truly led to a 
contradiction with semiclassical reasoning we might be forced to 
question it. However our point, made below, is that it {\it is} 
compatible with 
semiclassical reasoning, and hence there is no reason to suspect its 
validity.}. In such a theory black holes are effectively 
nonrelativistic, pointlike 
quantum particles. 
In the zero-charge sector of the Hilbert space, black holes are absent: 
The low-energy  effective field theory involves only the massless fields. 
However we may also consider the charged sector of 
the Hilbert space, for which the groundstate is in general degenerate 
and corresponds to a static collection of  
BPS-saturated black holes. Low energy 
excitations above such charged groundstates include non-relativistic motion 
of the black holes, and are also describable by 
effective field theory.  
In the D-brane phase this 
effective field theory contains one (super)field 
for every quantum state of the 
black hole:
\eqn\leff{{\cal S}_{eff}\sim-\sum_{i=1}^{e^{A/4}} 
\int d^4x ( (\nabla \Phi_i)^2 +m^2(\Phi_i)^2 + ...)}
Each field $\Phi_i$ creates a BPS-saturated supermultiplet. At low
energies above the groundstate, only non-relativistic modes of 
these fields contribute, and \leff\ reduces to the quantum mechanics
of slowly-moving black holes, as discussed below. The mass $m$
may depend on $g_s$ and other moduli but is highly constrained by 
supersymmetry. 
As mentioned above, in an $N=4$ or $N=8$ theory the number of 
such states typically does not change as a function of 
$g_s$. Hence \leff\ should be valid in the black hole phase as 
far as the number of fields. Determining all of the 
interactions however could involve solving an intractable strong-coupling 
problem.

It follows immediately from these assumptions, without any
knowledge
of the interactions, that the quantum state can be measured. Given 
that the states 
are described by an effective action of the form \leff, their identity
can be determined from 
interference experiments. For example in the scattering 
of two objects there is interference between Feynman diagrams which 
differ by the exchange of the final state objects if and only if they
are identical.   These experiments can be performed at 
low energies and large volumes, where effective field theory is valid.
It is also possible, by appropriate choice of the initial state, to avoid bringing the objects near to each other 
and the complicating effects of bound states at threshold. 
Repitition of such experiments will enable one to ascertain which 
of a large collection of like-charged black holes are identical and
which are not.

This conclusion might at first 
seem to be in conflict with semiclassical reasoning. 
Semiclassical low-energy scattering of two BPS black
holes 
with identical charges 
reduces to quantum mechanics on the two-black-hole
moduli 
space (after any high-energy modes of the $\Phi_i$ fields 
are integrated out). This
moduli space can apparently be reliably computed from the known classical 
two-black-hole solutions. It depends only on the charges and not on the 
quantum state of the black hole, which 
does not enter into the classical solutions. 
Hence the scattering would appear to be independent of the quantum state.

In fact there is an error in this reasoning. The classical moduli space is 
ambiguous due to a singularity on the subspace where the positions of the two 
black holes coincide. There is a $Z_2$ symmetry which exchanges the two
black holes and acts freely on the moduli space everywhere 
except at the coincident points. Because of this singularity 
one cannot determine from the 
classical solutions whether the moduli space is topologically 
${\cal M}_2 \sim R^3\times R^3$ or ${\cal M}_2 \sim (R^3\times R^3)/Z_2$. 
If we do divide by the $Z_2$,  the quantum mechanics will give  
interference, and otherwise not.  So the the semiclassical scattering in the 
black hole phase cannot be 
be unambiguously determined form low-energy field theory, and the 
possibility of 
interference is $\it not$ in conflict with low-energy reasoning. 

In string theory the moduli space can be unambiguously determined. At 
weak coupling the 
the D-brane picture tells us to divide by $Z_2$ if the black holes are
in the 
same 
state, and otherwise not. However the topology of the moduli space
cannot 
change 
as $g_s$ is varied, so this tells us that the moduli space should also be a 
$Z_2$ quotient in the black hole phase. This is somewhat reminiscent of 
Callan-Rubakov electron-monopole scattering: The low-energy effective 
theory has an ambiguity concerning the boundary condition at the origin,
which can only be resolved by consideration of the high-energy unified
gauge theory. 
More generally for a large collection 
of $N$ like-charged BPS black holes there is a discrete permutation 
symmetry $S_N$ which acts freely everywhere except at points
corresponding 
to one or more coincident 
black holes.  Hence classically the moduli space is determined only up to 
division by an arbitrary discrete subgroup of $S_N$,{\it i.e.} there is an 
ambiguity concerning whether or not the
exchange of two identically charged black holes returns one to the
same 
point in the multi-black hole moduli space.  All of these ambiguities
are 
resolved by going to weak coupling 
and using the D-brane picture, from which we are instructed to 
take the quotient by any 
element of $S_N$ which exchanges black holes in the same quantum
state. Once 
the identifications are determined, the moduli space is known and it
is possible to determine directly from the moduli space 
which black holes are in the 
same 
quantum state and which are not\foot{Further dependence of the moduli space 
on the quantum state arises when one considers the possible 
formation of bound states 
at threshold, which is proportional to wave 
function overlaps and hence suppressed at 
large volume. These bound states correspond to 
lower-dimensional branches of the 
moduli space which emanate from regions 
corresponding to one or more coincident 
black holes. There are many such branches labeled by the quantum microstate 
of the bound states. The connectivity of these various 
branches again depends on the microstate and not just the charges.}.

In conclusion, the quantum state of the black hole 
can be measured in the black hole phase. 
This does {\it not} contradict semiclassical reasoning 
because of an ambiguity in the semiclassical moduli 
space. Resolution of this ambiguity requires the data 
which determines the quantum state.

Some time ago it was observed that black holes can 
carry quantum hair which is measurable 
by interference experiments in which strings lasso black holes \qhair.
This 
indicated that 
classical solutions do not contain all the information about the 
low-energy quantum behavior of black holes.  Similarly here we see 
another example in which classical solutions do not
determine low-energy quantum behavior of black holes, leading to 
statistical quantum hair. 
Statistical hair was first considered in \gpp. 

In the present context, statistical 
hair arises from discrete gauge symmetries. 
When two identical D-branes
coincide, there is an enhanced $SU(2)$ gauge symmetry. 
The separation between the two 
D-branes corresponds to a vev for a scalar field in the adjoint of 
$SU(2)$ living on the D-branes. Exchanging the 
two D-branes corresponds to reversing the sign of the vev, 
which is a discrete $Z_2$ gauge transformation. More generally 
$S_N$ arises as the
Weyl group of an enhanced 
$SU(N)$ gauge symmetry. Hence the moduli space 
identifications arise from unbroken discrete gauge symmetries. 
Although the context is different, this suggests a connection 
with the discrete 
gauge hair discussed in \kw. In \cpw\ it was found that discrete gauge hair 
has exponentially small but measurable effects on the state of 
the quantum fields outside of the black hole. (Similar effects 
were found for axion hair in string theory in \fg.) It would be 
interesting to see if 
statistical hair also leads to such effects.

Some time ago it was speculated that black holes in string theory 
carry so much quantum hair that it is possible to completely
reconstruct the initial state which formed the black hole,
and that this might somehow lead to a resolution of 
the information puzzle \sch\ \cpw. One problem with this idea
was that, according to our understanding of string theory and quantum hair 
at the time, there seemed to be only a finite number of types of
quantum hair which could encode only a finite amount of information. 
However we now know that string theory has an arbitrarily large number 
of such unbroken discrete gauge symmetries, since there is one 
for every pair of 
D-branes and there can be arbitrarily large numbers of D-branes. 
Hence statistical hair can 
provide an arbitrarily large amount of information about a black hole, unlike 
previously discussed types of quantum hair. 
For this reason it is tempting to speculate that 
statistical hair may play a key role in the string theoretic 
resolution of the black hole information puzzle.

\vskip.2in

{\bf Acknowledgements}
I would like to thank T. Banks, S. Coleman, S. Giddings, 
J. Harvey, S. Hawking, G. Horowitz, R. Myers, J. Polchinski, 
L. Randall and N. Seiberg
  for useful 
discussions. I would like to thank the physics and mathematics 
departments at Harvard, the mathematics department at MIT, and 
the physics department at Rutgers for hospitality and support during
parts of this work.
This research of was supported in part by DOE grant DOE-91ER40618. 

\listrefs
\end